# Applications of Triggered Scaler Module for Accelerator Timing

Min Yang, Norihiko Kamikubota, Yuto Tajima, Kenichi C. Sato, Nobuhiro Kikuzawa

*Abstract*— Since 2006, there have been unexpected trigger-failure events during the operation of the Japan Proton Accelerator Research Complex (J-PARC) timing system. However, among the many suspected modules, it has been difficult to find the one with the fault. To find such a faulty module more easily, the triggered scaler module, which was designed as a PLC-type I/O module, was developed.

This type of module requires the start signal of the J-PARC main ring (MR) slow cycle (2.48/5.20 s) and the start signal of the rapid cycle (25 Hz) as reference signals, which are generated by the J-PARC timing system. A scaler in the module counts the number of input pulses during a machine cycle, and stores them in memory-buffers.

In 2018, the module was evaluated using the trigger for injection kicker and the low-level radio frequency (LLRF) signal of J-PARC MR. The expected results were shown. In 2020, two applications were developed based on a triggered scaler module. The first one, an application for detecting unexpected-trigger events, successfully detected simulated trigger-failure events. The second one, an application for the abort signal of a machine protection system (MPS), showed that the module is capable of visualizing signals based on a relationship with the accelerator cycle.

Two applications show possibilities for the development of more customized applications in the future. An idea to develop a portable unexpected-trigger detection system, which can be used in other accelerator facilities, is discussed herein.

## I. INTRODUCTION

The Japan Proton Accelerator Research Complex (J-PARC) is a high-intensity proton accelerator facility consisting of a 400-MeV linear accelerator (LINAC), a 3-GeV rapid-cycle synchrotron (RCS), a 30-GeV main ring (MR) synchrotron, and three research facilities, i.e., the Materials and Life Science Facility (MLF), Neutrino (NU) facility, and a Hadron (HD) facility. Using MW-class high-power proton beams, generated secondary particles are used in various advanced studies [1-3].

There are two time cycles used in J-PARC. A 25 Hz rapid cycle is used at the LINAC and RCS accelerators, and a slow cycle is used at the MR accelerator. When the MR delivers proton beams to the NU and HD facilities, 2.48 s (fast extraction mode (FX)) and 5.20 s (slow extraction mode (SX)) cycles are used, respectively. Because the slow cycle determines the overall time behaviour of the accelerators, it is also called a "machine cycle."

Fig. 1 shows one cycle of the MR timing scheme in SX mode (where the red line represents the time variation of the beam energy). The time marks from P1 to P4 indicate that the start of the MR injection, the start of the MR acceleration, the end of the MR acceleration and the start of the 30-GeV flat top, and the end of the flat top, respectively. These correspond to three main phases in one MR cycle: an injection phase (INJ), an acceleration phase (ACC), and a slow extraction phase (SX). During the injection phase, the MR accepts four successive injections in one machine cycle.

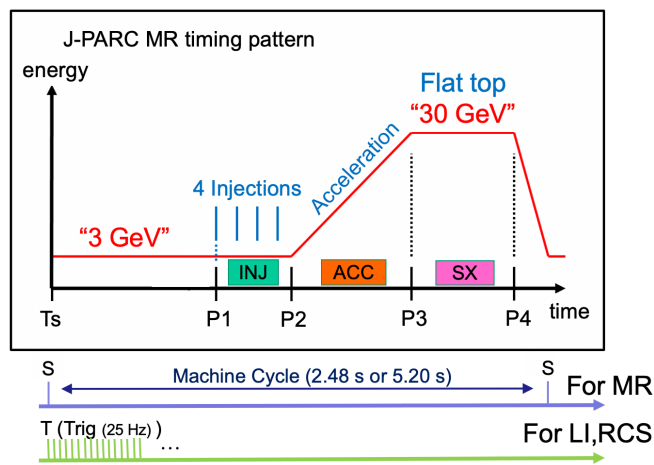

Fig. 1. Typical timing scheme of J-PARC MR.

The control system for J-PARC accelerators was developed using the Experimental Physics and Industrial Control System (EPICS) toolkit [4]. In addition, a dedicated timing system has been developed [5, 6]. The J-PARC timing system consists of one transmitter module and approximately 200 receiver modules. Both modules were developed as in-house VME modules. Event-codes, which are generated by the transmitter module, are distributed to the receiver modules. A fiber-optic cable network is used for event-code distribution using several optical-to-electrical (O/E) or electrical-to-optical (E/O) modules. According to the received event-code, each receiver module generates eight independent delayed trigger signals.

Since the first beam use began in 2006, the J-PARC timing

Min Yang is with the Graduate University for Advanced Studies (SOKENDAI) and the J-PARC Center, JAEA & KEK, 2-4 Shirakata, Tokai, Ibaraki 319-1195, Japan (e-mail: yangmin@post.kek.jp).

The work of Min Yang was supported by the China Scholarship Council (CSC).

Norihiko Kamikubota, Kenichi C. Sato, Nobuhiro Kikuzawa are with the J-PARC Center, JAEA & KEK, 2-4 Shirakata, Tokai, Ibaraki 319-1195, Japan (e-mail: norihiko.kamikubota@kek.jp).

Yuto Tajima is with the Kanto Information Service, 8-21 Bunkyo, Tsuchiura, Ibaraki 300-0045, Japan.

system has contributed to a stable operation of the accelerator beam [6]. Nevertheless, some timing trigger-failure events have occurred during beam operation. During each recovery process against a failure, it has often been difficult to find a definite module among the many modules suspected. Such experiences have prompted us to develop a new module that can read back signals generated by the J-PARC timing system. We developed a new module, called a triggered scaler module, for this purpose.

In this paper, the definitions of unexpected-trigger events and our experiences with them, the working principle and performance of the triggered scaler module, and the developments of two applications are described, followed by our future plans.

## II. Unexpected-Trigger Events

### A. Possible trigger-failure events

As a typical example, the trigger signal for the MR injection kicker is shown in Fig. 2 (a). During each machine cycle, the injection kicker injects beams four times to the MR. Therefore, there are four successive triggers.

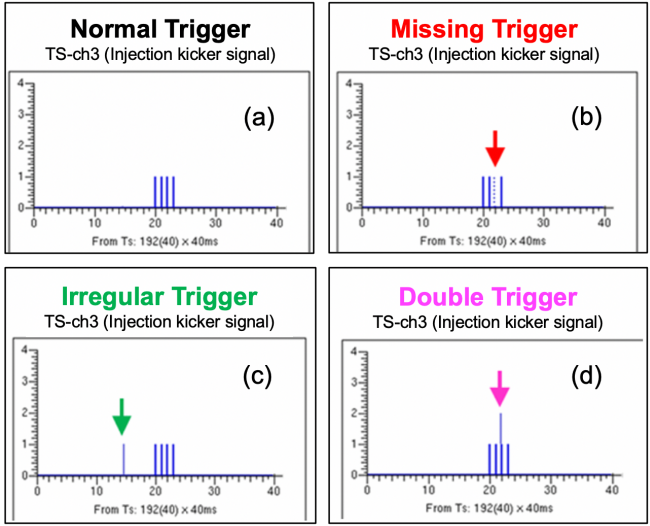

Fig. 2. (a) Normal signal of injection kicker trigger, (b) a missing trigger event, (c) an irregular trigger event, and (d) a double trigger event.

There are three possible failure events of the injection kicker trigger, as shown in Fig. 2 (b), (c), and (d). Fig. 2 (b) shows a missing trigger event, this means that one (or more) trigger disappears. Fig. 2 (c) shows an irregular trigger event, which indicates that an additional trigger is overlapped into the original signal. This event may be caused by noise. Fig. 2 (d) shows a double trigger event. This shows that one trigger is counted twice, which may be caused by a poor termination.

The trigger-failure events above are undesirable for an accelerator operation and are called "unexpected-trigger events."

### B. Previous unexpected-trigger events

Since 2006, we experienced some unexpected-trigger events during beam operation. Herein, two cases are given as examples, i.e., an irregular trigger event and a missing trigger event.

1) From November to December 2017, an O/E module, which was used to send a 25 Hz trigger signal from RCS to MR, started to produce irregular triggers (Fig. 3). Because the irregular triggers affected a critical beam diagnostic system, the accelerator operation was suspended several times per day [7]. It took 2 weeks to identify the troublesome O/E module.

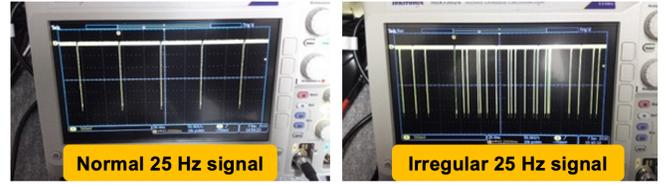

Fig. 3. The normal and irregular 25 Hz trigger signals monitored by an oscilloscope during beam operation.

2) A poor quality beam has rarely occurred during stable beam delivery to the HD facility. Here, "poor quality" means that the closed orbit distortion of the beam was slightly increased during the acceleration phase (Fig. 4). Such beams have appeared a few times per month since November, 2015 [7]. Finally, we found that a timing receiver module for one of the MR steering magnets showed momentary errors by external common-mode noises, resulting in a missing trigger. In May 2016, we added ferrite cores to metal cables connected to the receiver module as a countermeasure against the noise. This problem took approximately 6 months to solve.

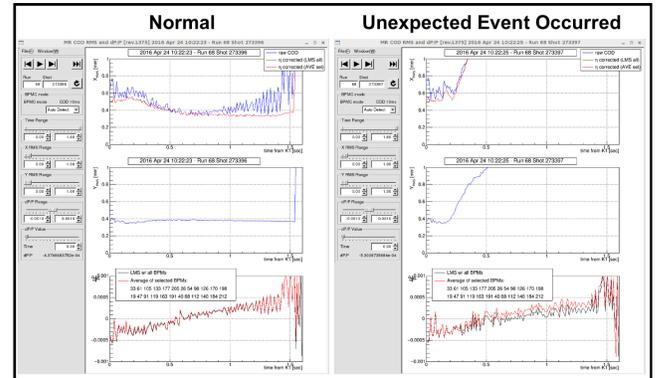

Fig. 4. Closed orbit distortions of MR: normal (left) and fatal (right).

The two examples above show that only one module, an O/E module or a receiver module, caused the unexpected-trigger events. Without a read-back system, it takes a long time to solve such a problem. To this end, a triggered scaler module was developed to read back signals of the J-PARC timing system.

## III. Triggered Scaler Module

### A. Overview

The triggered scaler (hereafter TS) module was designed as a key device for the timing read-back system. Because PLC is a standard I/O form in J-PARC MR, a Yokogawa PLC-type TS module was developed. An image and conceptual design of the module are shown in Fig. 5. Two types of reference signals are required: "S" (start signal of the machine cycle) and "Trig" (start signal of 25 Hz rapid cycle) signals, which are generated and distributed by the J-PARC timing system.

Before early 2018, four TS modules were produced as a prototype. One TS module has four input channels, and each channel has dual memory-buffers (16-bit, 192 cells $\times$ 2).

## B. Working principle

In principle, each channel of the TS module works as a scaler. Two inner FPGA logics are shown in Fig. 5. The logic (FPGA_1) counts the input pulses. When "S" arrives, FPGA_1 starts to increase the count in the first cell of the first memory-buffer. Each time "Trig" arrives, FPGA_1 shifts the pointer to the next cell. When the next "S" arrives, the pointer is moved to the first cell of the second memory-buffer. In other words, each cell keeps a number of trigger pulses received in a 40-ms bin. This scheme allows us to retrieve the number of counts per rapid-cycle during the last machine cycle. The other logic (FPGA_2) reads the memory-buffers, detects an unexpected-trigger in it, and sets error flags if necessary.

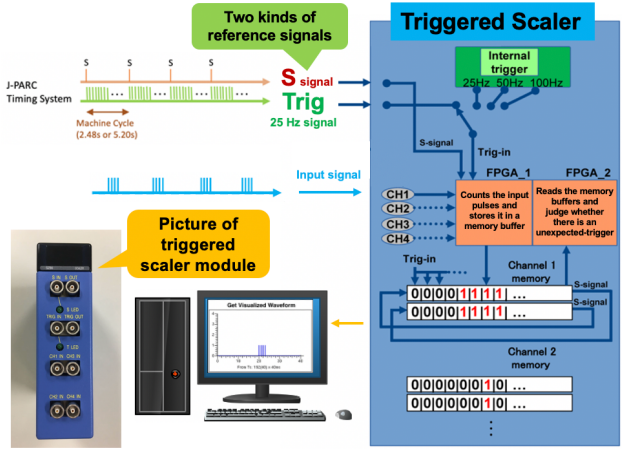

Fig. 5. Image and conceptual design of the triggered scaler module.

## C. Performance

To confirm the performance of the module, we prepared a test setup, as shown in Fig. 6. It consists of a CPU module, a TS module, and a power supply. Linux and EPICS are run inside the CPU module.

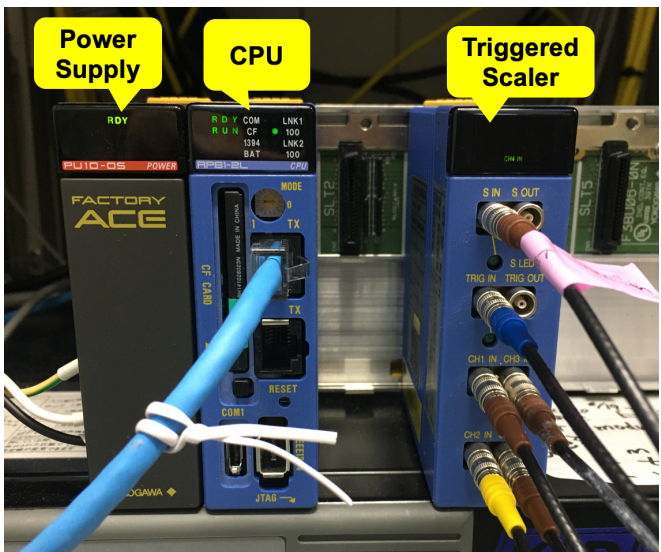

Fig. 6. Measurement setup of triggered scaler module.

In 2018, we measured a typical timing signal, i.e., the trigger signal for the MR injection kicker, which is shown in Fig. 2 (a). When the signal comes in, the scaler in the TS module counts the number of trigger pulses and stores them in the dual memory-buffers. The EPICS "caget" command prints out the memory-buffer data. As shown in Fig. 7, we successfully observed four successive "1" values in the memory-buffer of the module.

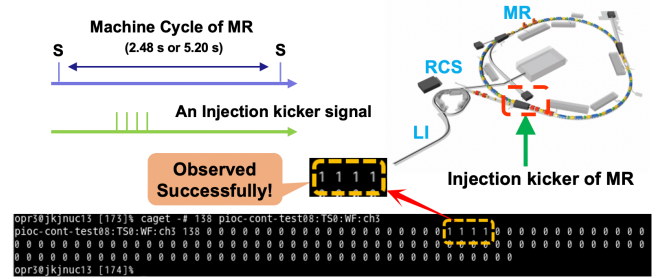

Fig. 7. Measurement of a trigger for the MR injection kicker.

Following the first measurement, we measured an RF signal (MR circulation signal), which was generated by a low-level radio frequency (LLRF) system. The 3-GeV proton beams are injected into the MR and then accelerated up to 30 GeV in each machine cycle. The graph in Fig. 8 shows a visualization of a frequency shift of an RF signal during the injection and acceleration phases. The number of counts in the 40-ms bin was measured using the TS. When the beam energy was 3 and 30 GeV, 7429 and 7647 counts were observed, respectively. As shown in Fig. 9, the observed counts are consistent with those expected based on the theory of relativity.

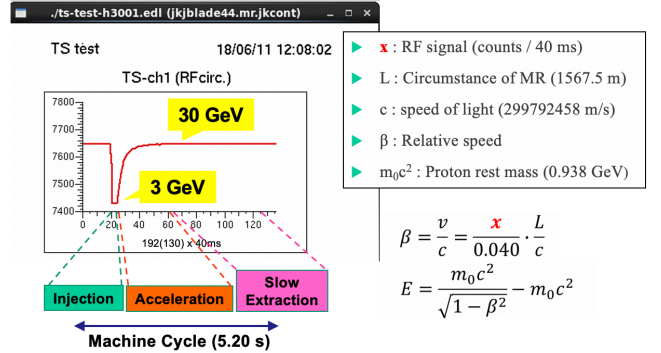

Fig. 8. Measurement of RF signal.

| | RF parameters | | Number of counts in 40-ms bin | |
|---|---|---|---|---|
| Energy (GeV) | RF (MHz)* | One turn (μs) | Expected | Observed |
| 30 | 1.7205 | 5.231 | 7647 | 7647 |
| 3 | 1.6717 | 5.384 | 7429 | 7429 |

Fig. 9. Number of counts in 40-ms bin, expected and observed, at two different energies.

The first measurement showed that the TS module operates as expected. It can be used to construct the read-back system of the J-PARC timing in the future. The second measurement showed that the TS can read back not only for the timing signals, but also for the LLRF signals. The two measurements demonstrated that the TS module is capable of detecting timing-related signals as well as visualizing them based on the relationship with the accelerator cycle.

## IV. APPLICATIONS BASED ON TRIGGERED SCALER MODULE

Using the same hardware setup shown in Fig. 6, two customized applications have been developed.

### A. Unexpected-trigger detection for injection kicker trigger

An application was developed to detect unexpected-trigger events of the injection kicker trigger. The EPICS database was developed to identify three types of unexpected-trigger events, as described in Fig. 2. It searches for non-zero or not-one values in the memory-buffer and identifies the type. An application GUI was developed using an EPICS tool.

First, we simulated three types of unexpected-trigger events using a dummy signal and tested the application. As shown in Fig. 10, the application successfully detected and identified all events.

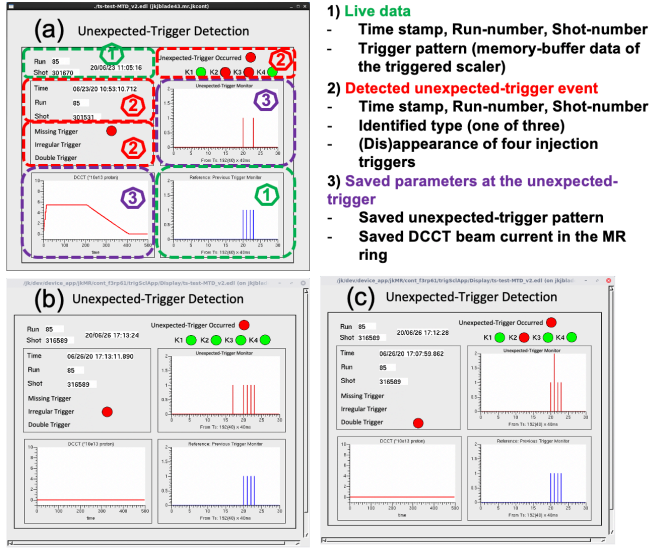

Fig. 10. Test results of the unexpected-trigger detection system using a dummy signal. Screenshots for (a) a missing trigger event, (b) an irregular trigger event, and (c) a double trigger event.

Second, the application was used with the real injection kicker signal during J-PARC beam operation. The result showed no unexpected-trigger event detection for approximately a 40-h period in June 2020.

### B. Machine Protection System - Abort Signal Detection

The machine protection system (MPS) is a fast interlock system that safely stops the accelerator operation. When one MR device, such as a magnet power supply or beam-loss monitor, detects a fault, the J-PARC MR-MPS system generates two signals. One is an "All-Stop" signal sent to the upstream LINAC to stop the beam operation, and the other is an "Abort" signal sent to the MR extraction kicker to abandon beams to the abort dump [8].

In the J-PARC MR, timestamps of MPS events are recorded by an archive system. To analyze an MPS event, we often need to know which machine phase of the MPS that occurred (see Fig. 1 for the machine phases). However, deriving this from the recorded timestamp is a complicated manual task.

Thus, we developed an MPS-abort detection application, which receives an abort signal from an MR-MPS unit and visualizes which machine phase the abort signal generated.

Fig. 11 shows a screenshot of the GUI application, in which the MPS event on June 18, 2020 is shown. The pointer position to the memory-buffer of the TS is shown as the "time-index." When the MPS occurred, it stopped at the 118th position, which corresponds to the SX phase. This fact is clearly visualized in Fig. 11. In addition, the DCCT beam current at the MPS was saved and indicated as a reference.

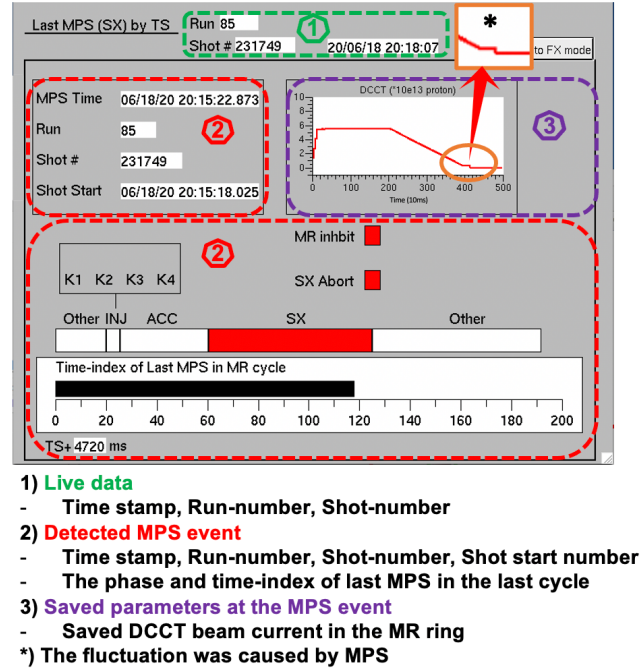

Fig. 11. Screenshot of MPS-abort detection system.

### C. Discussion

The first application, an unexpected trigger detection application, is an experimental product along the lines of the early motivation of TS development. The application is capable of detecting and identifying unexpected-trigger events, but only for the injection kicker trigger. More improvements allowing general trigger signals to be accepted have been discussed and will be achieved in the future.

The second application, an MPS-abort detection application, shows that a TS can be used to detect non-timing signals. We expect that this application will be used in beam operations after December 2020.

The experiences above show that, with the use of the same hardware setup, different applications can be customized by software, mainly EPICS database configurations. More ideas in the future will result in a wider use of TS modules in accelerator facilities.

## V. FUTURE PLANS

### A. Improvement of triggered scaler module

Two TS functions, the FPGA_2 logic and the internal trigger, shown in Fig. 5, have been designed but do not work well yet. Both will be evaluated after upcoming improvements in the firmware.

### B. Applications for rapid-cycle machines

To date, the developed applications have only been for J-PARC MR, which is a slow-cycle machine. We are currently developing applications for rapid-cycle machines, namely, the J-PARC LINAC.

### C. Portable unexpected-trigger detection system for other accelerators

The present setup assumes the use of the resources of the J-PARC control system. Thus, it is unavailable elsewhere. However, it is easy to configure the PLC-based module to be a stand-alone system. Considering the benefit of the small form-factor of the PLC, our system can be brought to other accelerator facilities.

Fig. 12 shows a possible portable system for an unexpected-trigger detection. It has additional I/O modules (a digital output and an ADC module), which apply specific actions when a fault event is detected by the TS. Actions are customized by updating the EPICS IOC software upon request. In addition, introducing a micro-server [9] as an archiver would be a powerful way to allow a later analysis of faulty events.

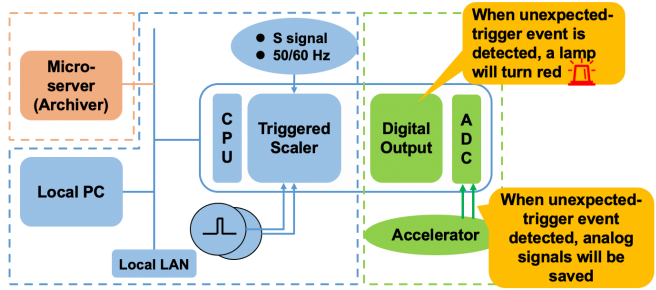

Fig. 12. Portable unexpected-trigger detection system with additional I/O modules and a micro-server.

## VI. CONCLUSION

A triggered scaler module was developed to obtain the read back signals of the J-PARC timing system. The module was evaluated using the trigger signal for the injection kicker and an RF signal. The results show that the module is capable of detecting timing-related signals as well as visualizing them based on the relationship with the accelerator cycle.

Using the module, two applications have been developed: an unexpected-trigger detection application and an MPS-abort detection application. The former successfully detects and identifies simulated unexpected triggers, but only for the trigger signal for the injection kicker. Improvements in accepting general trigger signals will be made. The latter shows the possibility of using the module for non-timing signals with software customization. More ideas for the wider use of the module are expected.

An idea to develop a portable unexpected-trigger detection system has been discussed. With additional I/O modules and a micro-server, the system will be expected to be used in other accelerator facilities.

## ACKNOWLEDGMENT

We express our best thanks to Noboru Yamamoto and other J-PARC accelerator staff members for their encouragement and suggestions during system development.